\documentclass{elsart3}

\usepackage{amssymb,amsmath}

\usepackage[english,francais]{babel}

\newtheorem{theorem}{Theorem}

\newtheorem{e-proposition}{Proposition}

\newtheorem{e-definition}{Definition\rm}

\def\og{\leavevmode\raise.3ex\hbox{$\scriptscriptstyle\langle\!\langle$~}}
\def\fg{\leavevmode\raise.3ex\hbox{~$\!\scriptscriptstyle\,\rangle\!\rangle$}}

\newcommand\R{\mathbb R}

\newcommand\Z{\mathbb Z}

\newcommand\eps{\varepsilon}
\newcommand{\vrho}{\varrho}
\newcommand{\la}{\langle}
\newcommand{\ra}{\rangle}

\renewcommand\P{\mathbb P}

\newcommand{\scal}[1]{\la #1 \ra}

\begin{document}
\centerline{}
\begin{frontmatter}


\selectlanguage{english}
\title{On localization for the Schr\"odinger operator with a Poisson random potential}


\selectlanguage{english}
\author[authorlabel1]{Francois Germinet\thanksref{FG}},
\ead{germinet@math.u-cergy.fr}
\author[authorlabel2]{Peter Hislop\thanksref{PH}},
\ead{hislop@ms.uky.edu}
\author[authorlabel3]{Abel Klein\thanksref{AK}}
\ead{aklein@math.uci.edu}

\address[authorlabel1]{D\'epartement de Math\'ematiques, Universit\'e de Cergy-Pontoise,
2 av. A. Chauvin,
95302 Cergy-Pontoise Cedex, France}
\address[authorlabel2]{Department of Mathematics, University of Kentucky, Lexington, KY 40506-0027, USA}
\address[authorlabel3]{University of California, Irvine,
Department of Mathematics,
Irvine, CA 92697-3875,  USA}

\thanks[FG]{Currently visiting the Universit\'e de Paris Nord with support from the CNRS.}
\thanks[PH]{Partially supported   by NSF Grant
DMS-0202656.}
\thanks[AK]{Partially supported   by NSF Grant
DMS-0200710.}


\begin{abstract}
\selectlanguage{english}
We prove exponential localization for the Schr\"odinger operator with a  Poisson random potential at the bottom of the spectrum in any dimension. We also prove exponential localization in a prescribed interval for all large Poisson densities. 
In addition,  we obtain  dynamical localization and finite multiplicity of the eigenvalues.

\vskip 0.5\baselineskip

\selectlanguage{francais}
\noindent{\bf R\'esum\'e} \vskip 0.5\baselineskip \noindent
{\bf }
On d\'emontre localization exponentielle pour l'op\'erateur de Schr\"odinger avec un potentiel al\'eatoire de Poisson, pour les basses energies et en toute dimension. On d\'emontre aussi localization exponentielle dans un intervalle d'\'energies donn\'e et \`a grande densit\'e. On obtient de plus localisation dynamique et le fait que la multiplicit\'e des valeurs propres est finie.

\end{abstract}
\end{frontmatter}

\section{Results}

The Poisson Hamiltonian is  the random Schr\"odinger operator on $\mathrm{L}^2(\mathbb{R}^d)$
given by 
\begin{equation}
H_X =  -\Delta + V_{X} , \quad \text{with} \quad  V_{X}(x)=  \sum_{\zeta \in X} u(x - \zeta),
\end{equation}
where the single-site potential  $u$ is a  nonnegative  $C^{1}$ function on $\R^{d}$ with compact support--without loss of generality we take $u(x) \le u(0)=1$--and  $ V_{X}$ is a Poisson random potential, that is,  $X$ is a Poisson process on $\R^d$ with density $\varrho >0$.  Thus the configuration   $X$ is a random countable subset of $\R^{d}$, and, letting $N_{X}(A)$ denote the number of points of $X$ in the Borel set $A\subset \R^{d}$, each $N_{X}(A)$ is a Poisson  random variable with mean  $\varrho |A|$  (i.e., $\P_\varrho\{N_{X}(A)=k\}=(\varrho |A|)^k (k!)^{-1} \mathrm{e}^{-\varrho |A|}$ for $k=0,1,2,\dots$), and the random variables  $\{N_{X}(A_{j})\}_{j=1}^{n}$ are independent for disjoint Borel sets $\{A_{j}\}_{j=1}^{n}$.  We will denote by $(\mathcal{X},\P_{\vrho})$  the underlying probability space for the Poisson process with density $\varrho $.

Note that $H_{X}$ is an ergodic (with respect to translations in $\R^{d}$) random self-adjoint  operator. It follows that the spectrum
of $H_X$ is the same for $\P_\varrho$-a.e. $X$, as well as the decomposition of the spectrum into pure point, absolutely continuous, and singular continuous spectra.
For $u$ as above we actually get $\sigma(H_X)=[0,+\infty[$ for $\P_{\vrho}$-a.e. $X$  \cite{KM}.

We prove exponential localization  for Poisson Hamiltonians at the bottom of the spectrum. By  $\chi_B$ we denote the characteristic  function of the
set $B \subset  \mathbb{R}^d$, with 
 $\chi_{x}$  denoting the characteristic  function of the
 cube of side $1$ centered at
 $x \in \mathbb{R}^d$. We write
$\langle x \rangle= \sqrt{1+|x|^2}$, 
$T(x)=\scal{x}^\nu$ for some fixed $\nu>\frac d 2$. 

\begin{theorem}\label{thmpoisson}
Given $\varrho>0$, there exists $E_{0}=E_0(\vrho)>0$ and $m= m(\rho)>0$, such that for $\P_\varrho$-a.e. $X$ the following holds: the operator $H_X$ has pure point spectrum  in $[0,E_0]$ with exponentially localized eigenfunctions with rate of decay $m$, i.e., if    $\phi$ is an eigenfunction of $H_X$ 
with eigenvalue $E \in[0,E_0]$,  there is  a constant $C_\phi<\infty$ such that 
\begin{equation}\label{expdecay}
 \|\chi_x \phi\| \le C_\phi e^{-m|x|} \quad \text{for all $x \in \R^{d}$}.
\end{equation}
Moreover there exist constants  $\tau>1$, $s\in]0,1[$, and  $C<\infty$, such that for   eigenfunctions $\psi,\phi$ (possibly equal) with eigenvalue  $E\in[0,E_0]$ 
we have 
\begin{equation}\label{SUDEC}
\| \chi_x\psi\| \, \|\chi_y \phi\| \le C\|T^{-1}\psi\|\|T^{-1}\phi\| \, e^{\scal{y}^\tau} e^{-|x-y|^s} \quad \text{for all $x,y \in \Z^{d}$}.
\end{equation}
 In particular, the eigenvalues of $H_X$ in $[0,E_0]$ have finite multiplicity,  and  $H_X$ exhibits dynamical localization in $[0,E_0]$, that is, for any $p>0$  we have
\begin{equation}\label{dynloc}
\sup_t \| \scal{x}^p e^{-itH_X} \chi_{[0,E_0]}(H_X) \chi_0 \|^2_2 < \infty.
 \end{equation}
\end{theorem}

For Poisson random potentials the density $\vrho$ is a measure of the amount of disorder in the medium.  The next theorem gives localization at high disorder.

\begin{theorem}\label{thmpoissonbis}
Given $E_0 >0$, there exists $\varrho_0>0$  such that for $\varrho>\varrho_0$ the conclusions of Theorem~\ref{thmpoisson} hold in the interval  $[0,E_0]$.
\end{theorem}

While Poisson Hamiltonians  are the most natural random Schr\"odinger operators  in the continuum (the distribution of impurities in a material being naturally modeled by a Poisson process), a mathematical proof of the existence of localization has been a long-standing open problem.  Localization has been known only in one dimension \cite{St}. A Poissonian model, which incorporates random intensities with bounded densities and requires single-site potentials that do not decay too slowly at infinity, was considered in \cite{CH}.

 In the multi-dimensional case,  localization in the continuum had been proved  for Anderson-type Hamiltonians with random intensities with bounded densities, e.g.  \cite{CH}, and  for an $\R^d$-ergodic Schr\"odinger operator with a Gaussian random potential  \cite{FLM}; in both cases there is an ``a priori" Wegner estimate obtained by averaging with bounded densities.   But recently   Bourgain and Kenig  \cite{BK} proved localization for the Bernoulli-Anderson Hamiltonian, with the Wegner estimate being proven in a multiscale analysis.

To prove Theorems~\ref{thmpoisson} and \ref{thmpoissonbis} we exploit the  new ideas introduced by  Bourgain and Kenig  \cite{B,BK}. In particular, the control of the resonances (the Wegner estimate) is achieved by a multiscale analysis using  ``free sites" and a new quantitative version of unique  continuation which gives  a lower bound on eigenfunctions.

The control on the eigenfunction correlations given in (\ref{SUDEC}) was introduced in \cite{GK2}. That (\ref{SUDEC}) implies dynamical localization is rather immediate. As for the finite multiplicity property, it follows by estimating $\| \chi_x \chi_{\{E\}}(H_X)\|_2^2 \|\chi_y  \chi_{\{E\}}(H_X)\|_2^2$ from (\ref{SUDEC}) and summing over $x \in \Z^d$.

In the next section we outline the main ideas in the proof of Theorem~\ref{thmpoisson} ;  detailed arguments will be given in  \cite{GHK}.
Theorem~\ref{thmpoissonbis} is proved in a similar way, although the proof requires some modifications.

\section{The main ideas}

  Given   a cube $\Lambda$  in $\R^{d}$, we let 
$H_{X,\Lambda}=-{\Delta_{\Lambda}}+ V_{X,\Lambda} $ be the restriction of $H_{X}$ to $\Lambda$ with Dirichlet boundary condition.
We consider the finite volume resolvent 
 $R_{\Lambda} (E)=(H_{\Lambda} - E)^{-1}$ (we will omit $X$ from the notation).  By  $c,c^{\prime}, \dots$ we  denote positive constants (not necessarly the same) independent of $\vrho, L, \dots$.

The multiscale analysis requires  an {\it a priori} probabilistic  estimate on the resolvent $R_\Lambda (E)$ for all $E \in [0,E_{0}]$ at a fixed, but sufficiently large, initial scale $L_{0}$, where $\Lambda$ is a cube of side $L_{0}$ centered 
at, say, $x_0$.
To obtain this  initial estimate for  Theorem~\ref{thmpoisson}, we divide the cube  $\Lambda$ into non-overlapping cubes $\Lambda(j)$ of side
 $\ell_{0} \approx ( { \vrho}^{-1} \log L_{0})^{\frac1d}$  centered at points 
 $j \in x_{0} +          {\ell_{0}} \Z^{d}$.  We consider configurations such that
  $1\le N(\Lambda(j)) \lesssim \vrho         {\ell_{0}}^{d}$ for all
$\Lambda({j})$, an event with high   probability, more precisely,
with probability   $\ge 1 -\left( \frac {L_{0}}          {\ell_{0}}\right)^{d} L_{0}^{-p}
  \ge 1 -  \vrho L_{0}^{-p+d}$, where we can arrange for  $p$  large as desired.

For such configurations, we pick one  $\zeta_{j}  \in \Lambda({j})$, and split the potential as $V_{\Lambda}= V_{\Lambda}^{(1)} + V_{\Lambda}^{(2)}$, with 
 $ V_{\Lambda}^{(1)}=  \sum^\prime_j u(x - \zeta_{j})$, where $\sum^\prime_j$ denotes the sum over  sites ${j\in x_{0} + 2         {\ell_{0}} \Z^{d}}$ only; as a consequence the $ u(x - \zeta_{j})$ in the sum are non-overlapping.    We have  $0\le V_{\Lambda}^{(1)}\le 1$ and $V_{\Lambda}^{(2)}\ge 0$. In order to estimate
$\| R_{\Lambda}(E) \|$, it is convenient to use the
 operator $\Gamma_\Lambda (E)$,  defined by
\begin{equation}\label{BS}
\Gamma_\Lambda (E) = (H_\Lambda^{(2)}+1)^{-\frac12} (1+E-V_\Lambda^{(1)}) (H_\Lambda^{(2)}+1)^{-\frac12},\quad \text{with $H_\Lambda^{(2)}=-\Delta_{\Lambda } +V_{\Lambda}^{(2)} \ge 0$}.
\end{equation}
Proceeding as in \cite[Section~4]{BK}, suppose $\|\Gamma_\Lambda (E) \|> 1 -E_{0}$ with $E_{0}$ small. Then there is  $g\in\mathrm{L}^2(\R^d)$, with
 $1 -\sqrt{E_{0}}\le \|g\|\le 1$ and, using $ V_{\Lambda}^{(2)}\ge 0$ , $ \| \nabla g \| \le 2E_0^{\frac14}$,  such that for each $a \in \Lambda$ we have
 \begin{equation}\label{translation}
0 \le \langle \tau_a V_\Lambda^{(1)} g, g \rangle \le c E_0^{\frac14}
(|a|+1),
\end{equation}
where $\tau_{a}$ denotes translation by $a$ and and the estimate is uniform in $L$
(cf.  \cite[Eqs. (4.7),  (4.8), and (4.10)]{BK}).
On the other hand, taking $K=10  {\ell_{0}}$,  and recalling the definition of
$V_{\Lambda}^{(1)} $, we get (cf. \cite[Eqs. (4.12) and (4.15)]{BK})
 \begin{equation}\label{average}
 \int_{[-K,K]^{d}}  \tau_a (V_\Lambda^{(1)}) \mathrm{d} a
\ge c \chi_{\Lambda_L} \quad \text{with $c >0$}.
\end{equation}
Combining (\ref{translation}), (\ref{average}), and the lower bound on $\|g\|$, we get    $c(1 -\sqrt{E_{0}})^{2 } \le c^{\prime} E_0^{\frac14} K^{{d+1}}$,
which leads to a contradiction for $E_0 \approx   {\ell_{0}}^{-(4(d+1)+)}$ and $L_0$ large.

We may thus conclude that if $\vrho$ is fixed, $p>0$ is given,  $E \in [0,E_{0}]$ with  $E_{0} \approx  ( { \vrho}^{{-1}} \log L_{0})^{-(\frac {4 (d+1)}d   +)}$, and  $L_0$ is  sufficiently large,  then,  with probability   $  \ge 1 -  L_{0}^{-p}$, we have  $\|R_{\Lambda }(E)\|\lesssim E_{0}^{-1} $ and $\| \chi_{x} R_{\Lambda }(E) \chi_{y}\| \lesssim \e^{-c L_{0}} $ for
$x,y \in \Lambda$ with $ |x-y| \ge \frac {L_{0}}{10}$.   Moreover, it is clear that if 
 $V_{\Lambda}^{(2)}=   \sum_{{\zeta \in Y}} u(x - \zeta)$, the results are still valid if we replace
 $V_{\Lambda}^{(2)}$ by  $ \sum_{{\zeta \in Y}} t_{\zeta }u(x - \zeta)$ with  arbitrary
 $t_{\zeta} \in [0,1]$.  We now declare all boxes $\Lambda(j)$ with ${j \notin x_{0} + 2         {\ell_{0}} \Z^{d}}$  (and hence do not contribute
 to $V_{\Lambda}^{(1)}$) to be \emph{free boxes}. Moreover, inside the free boxes we use the representation of the Poisson process $X$ by a thinned Poisson process (e.g., \cite{Reiss}), that is, by a Poisson process $Y$  with density $2\vrho$ in such a way that to each Poisson point $\xi \in Y$ is attached a Bernoulli random variable $\eps_{\xi}$, $\eps_{\xi}=0$ or $1$ with equal probability, and the single-site contribution to the potential is given by $\eps_{\xi} u(x-\xi)$.  Note that any site $\xi \in Y$ in a free box is a \emph{free site} in the sense of \cite{BK}.
 
The multiscale analysis now proceeds by induction. If $\Lambda$ is a box of size $L$, we divide it into non-overlapping cubes $\Lambda(w)$ of side $\approx \e^{-L^{2}}$ centered at points $w \in \e^{-L^{2}}\Z^{d}$.  With  probability $\ge 1 - L^{-p}$,  $p$ large, we require $N(\Lambda) \lesssim \vrho L^{d}$ and
all  $N(\Lambda(w)) \le 1$.  We introduce an equivalence relation on
Poisson configurations $X_{\Lambda}$ in $\Lambda$; 
$\widetilde{X}_{\Lambda}$ is the collection of Poisson configurations in $\Lambda$ that cannot be distinguished from $X_\Lambda$ by the counting functions  $N(\Lambda(w))$.  The crucial observation is that if we change
a Poisson configuration to another one  in the same equivalence class, then 
the eigenvalues of $H_{\Lambda}$ in a fixed interval do not move by more than
$\lesssim  \e^{-L^{2}}$. We may thus consider only the case when the Poisson points in $\Lambda$ are in the lattice  $\e^{-L^{2}}\Z^{d}$, since the desired results will then hold for the whole equivalence class.  This reduction allows the use of the results in \cite{BK}, using equivalence classes of Poisson configurations instead of fixed Bernoulli configurations. Inside the free boxes equivalence classes are defined
as above but for the Poisson points in $Y$. Since we have a finite  number
of equivalence classes of configurations inside a free box, we fix the points of the Poisson process $Y$ in the free boxes, and conduct the analysis of \cite[Lemmas 5.1]{BK},   ``tuning'' the free parameters $t_\xi$ to $\eps_\xi=0$ or $1$ to obtain ``good''  configurations, with a probability estimated
by Sperner's Lemma using  \cite[Lemma 3.1]{BK}. As in \cite{BK}, we get the following result (cf. \cite[Proposition A]{BK}), where $\Lambda_{L}$ denotes a cube of side $L$.

\begin{e-proposition}\label{propMSA} Given $\varrho>0$, there exists $E_{0}=E_0(\vrho)>0$ and $L_{0}=  L_{0}(\varrho) < \infty$, such that
if $\mathcal{X}_{\Lambda_{L}}(E)$ denotes the Poisson  configurations
for which
\begin{equation}\label{good1}
\| R_{\Lambda_{L}}(E) \| \le e^{L^{1-}} \quad \text{and} \quad 
\| \chi_x R_{\Lambda_{L}}(E) \chi_y \| \le e^{-c L}\;  \text{for}\;  |x-y|\ge \frac L{10},
\end{equation}
then
for all $L\ge L_0$ and all $E \in [0,E_{0}]$ we have 
\begin{equation}\label{weakest}
\P_\varrho\left\{\mathcal{X}_{\Lambda_{L}}(E)\right\}
\ge 1 -  \frac1{L^{\frac38 d-}}.
\end{equation} 
\end{e-proposition}

Proposition~\ref{propMSA} provides a single-energy multiscale analysis.
The weak probability estimate  in \eqref{weakest}  does not allow for an energy-interval multiscale analysis  as in  \cite{vDK,GK1}.  The first part of Theorem~\ref{thmpoisson}, namely exponential localization, requires the energy elimination scheme given in \cite[Section 7]{BK}. To obtain  the decay of the eigenfunction correlations given in  \eqref{SUDEC} we add ideas from \cite{GK2}.


\end{document}